\documentclass[superscriptaddress,twocolumn,aps,pra,preprintnumbers,notitlepage,showkeys,showpacs]{revtex4-1}
\usepackage{amsmath,amssymb}
\usepackage{lipsum} 
\usepackage[utf8]{inputenc}
\usepackage{graphicx}
\usepackage{dcolumn}
\usepackage{bbold}
\usepackage{bm}
\usepackage[usenames]{xcolor}
\usepackage[colorlinks,bookmarks=false,citecolor=blue,linkcolor=blue,urlcolor=blue,hyperfootnotes=true]{hyperref}
\usepackage{accents}

\makeatletter
\usepackage{times}
\usepackage{lineno} 
\usepackage{dcolumn}
\usepackage{bm}
\usepackage{braket}
\newcommand{\beq}{\begin{equation}}
\newcommand{\eeq}{\end{equation}}
\newcommand{\bqa}{\begin{eqnarray}}
\newcommand{\eqa}{\end{eqnarray}}

\newsavebox{\mybox}

 \let\oldequation\equation
\let\oldendequation\endequation

\renewenvironment{equation}
  {\linenomathNonumbers\oldequation}
  {\oldendequation\endlinenomath}

\let\oldalign\align
\let\oldendalign\endalign

\renewenvironment{align}
  {\linenomathNonumbers\oldalign}
  {\oldendalign\endlinenomath}
 
\makeatother

\begin{document}
\title{Incoherent transport in a model for the strange metal phase: Memory-matrix
formalism}
\author{Emile Pangburn}
\affiliation{Institut de Physique Théorique, Université Paris-Saclay, CEA, CNRS,
F-91191 Gif-sur-Yvette, France}
\author{Anurag Banerjee}
\affiliation{Department of Physics, Ben-Gurion University of the Negev, Beer-Sheva 84105, Israel}
\author{Hermann Freire}
\affiliation{Instituto de {Fí­sica}, Universidade Federal de Goiás, 74.001-970,
Goiânia-GO, Brazil}
\author{Catherine Pépin}
\affiliation{Institut de Physique Théorique, Université Paris-Saclay, CEA, CNRS,
F-91191 Gif-sur-Yvette, France}
\begin{abstract}
We revisit a phenomenological model of fermions
coupled to fluctuating bosons that emerges from finite-momentum particle-particle pairs for describing the strange
metal phase in the cuprates. The incoherent bosons dominate the transport properties
for the resistivity and optical conductivity in the non-Fermi liquid phase. Within the Kubo formalism, the resistivity is approximately
linear in temperature with a Drude form for the optical conductivity, such that the Drude
lifetime is inversely proportional to the temperature. Here,
we compute the transport properties of such bosons within the memory-matrix approach that successfully captures the hydrodynamic regime. This
technique emerges as the appropriate framework for describing the transport
coefficients of the strange metal phase. Our analysis confirms
the $T$-linear resistivity due to the Umklapp scattering that we obtained for this effective model. Finally, we provide new predictions
regarding the variation of the thermal conductivity with temperature and examine the validity of the Wiedemann-Franz law. 
\end{abstract}
\maketitle

\section{Introduction\label{subsec:Introduction}}

One of the most enduring mysteries of quantum condensed matter physics
is arguably the strange metal phase of the cuprate superconductors
\cite{Keimer2015FromQM,Norman03,Lee06}. The
conventional metal obeys universal laws for the variation
of transport coefficients with temperature. {The standard transport theory of metals gives a simple dependence of the longitudinal $\sigma_{xx}$ and Hall $\sigma_{xy}$ conductivities as a function of transport lifetime with $ \sigma_{xx} = n e^2 \tau / m$ and  $\sigma_{xy} =n e^3 B \tau^2/c$, where $n$ is the number of electrons, $e$ is the elementary charge, $\tau$ is their lifetime, $m$ is their mass, $ B $ is the magnetic field and $c $ is the speed of light.} At low temperatures, the
inverse of the transport time typically goes like $\tau^{-1}\sim T^{2}$; hence, the longitudinal
conductivity displays $\sigma_{xx}\sim T^{-2}$, whereas the Hall
conductivity is given by $\sigma_{xy}\sim T^{-4}$ . Within the Fermi
liquid theory, which describes the behavior of conventional metals,
electronic quasiparticles are the sole type of charge carriers
and, therefore, the Hall angle becomes $\cot\theta_{H}=\sigma_{xx} /\sigma_{xy}\sim T^{2}$.

By contrast, the experimental data in the strange metal phase of the
cuprates display striking discrepancies with the standard Fermi liquid
picture~\cite{gurvitch1987resistivity,EmeryVJ:1995es,hussey2004universality}. Firstly, the experimental observations demonstrate that $\sigma_{xx}\sim T^{-1}$
\cite{Legros_2018} and, at the same time, $\cot\theta_{H}\sim T^{2}$
\cite{clayhold1989hall,Ando_Murayama,barivsic19}. Therefore, the transport time induced
from the longitudinal conductivity scales as $\tau\sim T^{-1}$, while
the ``Hall lifetime'' varies as $\tau_{H}\sim T^{-2}$, which is commonly referred to as the ``separation of lifetimes''
in the literature \cite{Coleman_1996,Varma89,Varma_MRL_hall}. Furthermore, the Wiedemann-Franz law is {satisfied (with a doping-dependent overall coefficient)} in 
the strange metal phase almost down to $T=0$~\cite{Cyril02_PRL,WFPRX18,WFPRB16}.
The notable disagreement between the different experimental data with the
Fermi-liquid paradigm makes the strange
metal phase of the cuprates one of the biggest enigmas of correlated 
quantum matter~\cite{Keimer2015FromQM,Varma89,Patel-PRB,patel2018magnetotransport,zaanen2019planckian}.

{The theoretical concepts which have been put forward to explain this very unusual situation can be summarized as follows. Quantum critical theories based on fermions interacting with Landau-damped critical bosons have been widely studied~\cite{Abanov00}. In these scenarios, electric charge is solely carried by the fermions. From this perspective, two cases emerge. First, suppose the bosons have finite momentum like in the antiferromagnetic quantum critical theory, among others. In that case, the fermions around the Fermi surface are partially sensitive to the scattering via the bosons. The proportion of the fermions that participate in such scattering is called ``hot" fermions, and the remaining part of the Fermi surface remains insensitive to the critical bosons. Since the latter fermions do not participate in the scattering, they are referred to as ``cold" fermions. This scenario was first identified in a seminal paper by Hlubina and Rice \cite{hlubina1995resistivity}. This mechanism is the principal obstacle to obtaining a linear-in-$T$ resistivity in these models within the clean limit. Indeed, at low enough temperatures, the ``cold" fermions short-circuit the ``hot" ones, and the transport lifetime falls back into the standard Fermi liquid paradigm with $\tau^{-1} \sim T^2$ (see also Ref. \cite{RoschPRL} for an example of this mechanism at play in the context of such systems with the subsequent addition of disorder).} 

{The second possibility within the fermion-boson quantum critical scenario is that the whole Fermi surface becomes ``hot", namely, that the critical bosons have zero momentum so that every fermion at the Fermi surface can participate in the scattering via the bosons (like in the Ising-nematic quantum critical theory, among others). Here all the fermions are ``hot"; there is no issue with a possible ``short-circuiting" with cold species. A caveat with this scenario is that a $T$-linear resistivity needs to be clearly obtained within the clean limit, with the temperature dependence of the resistivity varying from sublinear at high temperatures to quadratic at low enough temperatures~\cite{Berg2019}. Altogether, accounting for such a linear-in-$T$ resistivity with only fermions as charge carriers is challenging. New proposals then emerged, introducing new strongly coupled fixed point models within the Planckian limit of dissipation~\cite{zaanen2019planckian}. These very innovative scenarios (e.g.,~\cite{Patel-PRB}) have in common that charge carriers are not well-defined and that the systems are analogous to a highly-correlated plasma carrying the current. Strongly correlated fixed point models, including, e.g., the Sachdev-Ye-Kitaev (SYK) model (see~\cite{Chowdhury_RMP}), obtain the linear-in-$T$ behavior in the resistivity very elegantly. However, it is not clear yet how to deal with the ``two-lifetime" problem within such a scenario. Also, the discussion of the fundamental Planckian limit in these models led to interesting holographic descriptions using hydrodynamic modes and symmetries (see, e.g., \cite{Delacretaz2017}).}

In the present manuscript, we
 address such a longstanding issue by proposing a new type of bosonic excitation that can
potentially describe the strange metal phase in the cuprates \cite{putzke2019reduced,barivsic19,Bozovic04,Caprara17,Efetov13,Kontani:2008fd,McKenzieDMFT00}. 
{We would like to stress that the  scenario explored below  presents a unique case where the whole picture, including linear-in-$T$ 
resistivity and the cotangent of the Hall angle, is addressed, and contact with experiments is made possible}.
The physical picture underlying our phenomenological
model \cite{Banerjee20,pepin2022charge} can be summarized as follows:
At high energies, the microscopic lattice model generates fluctuating finite momenta bosons created
from particle-particle pairs (remnants of a pair-density-wave) and the constituent fermions. These
fluctuating bosons carry an electric charge of $2e$ and, hence, also contribute to the charge transport. 
Moreover, due to the scattering with fermions, such bosons become
incoherent as the bosonic propagator becomes $D^{-1}(\mathbf{q},\omega)=-i\omega+\mathbf{q}^{2}+m_{b}(T)$
(where $m_{b}(T)$ is the temperature-dependent bosonic mass). 
Such bosons in two dimensions indeed lead to a $T$-linear contribution
to the longitudinal conductivity 
and also the optical conductivity attains a Drude form $\sigma_{xx}^{\rm boson}(\omega)\sim(\tau_{\rm boson}^{-1}-i\omega)^{-1}$, as
shown in Ref.~\cite{Banerjee20}. Interestingly, recent study~\cite{Yang2022} reveals a bosonic strange metal phase in nanopatterned YBCO samples. In that work, the longitudinal conductivity shows a linear temperature and field dependence along with a vanishing Hall coefficient, as soon as
the bosonic transport sets in~\cite{Yang2022}.  {Moreover, we also point out the Ref.~\cite{putzke2019reduced}, 
where another charged carrier is suggested besides the fermions. This additional charge carrier has the experimental signature
 of contributing to the linear-in-$T$ behavior in the longitudinal resistivity, whereas it does not contribute to the Hall conductivity.}

Indeed, since the critical bosons {turn out to be particle-hole symmetric}, they do not contribute
to the transverse Hall conductivity $\sigma_{xy}^{\rm boson}=0$. Such
a general picture could explain the experimental data if the bosons are light enough to short-circuit the fermions for the longitudinal
conductivity. Since the fermions are also present in this model, there must be scattering between these two excitations.
{As stated above, at low temperatures, }scattering via incoherent bosons with finite
momentum produces a finite lifetime for the fermions, at least
on parts of the Fermi surface, with the generation of ``hot spots'' where the
dominant scattering is through the incoherent bosons, leading to a
scattering rate $\tau_{hot}^{-1}\sim T^{\alpha}$, with $1<\alpha<1.5$ \cite{hlubina1995resistivity,RoschPRL}. {As explained before, the part of the Fermi surface which is unaffected by the boson scattering is called ``cold".}
The modeling of the Hall conductivity on a Fermi surface with an anisotropic
lifetime has been treated in another study to describe the strange metal phase of the cuprates \cite{Hussey_McKenzie}.
The angular average on the Fermi
surface favors \cite{Hussey_McKenzie} the ``hot regions'' for the Hall conductivity, leading
to an average Hall inverse lifetime $\tau_{H}^{-1}\sim T^{3/2}$.
Our study thus combines the two types of excitations (bosons
and fermions) to give a new perspective to the old paradox, such that the Hall angle
becomes $\cot\theta_H\sim T^{3}/T\sim T^{2}$, consistent with the experiments.

{ Since the model of fermion-boson ``soup" with charged-two bosons is one of the few proposals for a regime with linear-in-$T$ resistivity and $ \cot \theta_H\sim T^2$,  and considering the very scarce number of studies of transport due to  charged bosons,  it is important to check the universality of this regime and to use another approach to calculate transport properties instead of the Kubo formula implemented in Ref. \cite{Banerjee20}. } In the present paper, we revisit this problem  { in the context of the hydrodynamic description used  in the discussion of the Planckian regime \cite{Sachdev-MIT,Delacretaz2017}} and confirm our key results
that such charge-two incoherent bosons in two dimensions contribute to the longitudinal
conductivity as $\sigma_{xx}\sim T^{-1}$. To this end, we investigate the transport properties
of such bosons using the memory-matrix technique, that successfully captures
the hydrodynamic regime. Consequently, the $T$-linear resistivity regime of the  {Landau-damped} charged bosons with finite momentum due to Umklapp interactions stands on firm ground. Finally,
we provide new predictions regarding thermal conductivity as a function of temperature in the
model and also discuss the validity of the Wiedemann-Franz law for this system.

\section{The model \label{subsec:The-model}}

We consider here a two-dimensional phenomenological model~\cite{Banerjee20}
of fluctuating charge-two bosons interacting with each other and among themselves. The
bosons are in a ``soup'' of fermions, and
the corresponding fermion-boson scattering affects the boson lifetime significantly (to be explained below).
The bosonic part of the Hamiltonian is given by
\begin{align}
\hat{\mathcal{H}} & =\sum_{\mathbf{q}}b_{\mathbf{q}}^{\dagger}\left[\frac{\text{\ensuremath{\left|\mathbf{q}\right|}}^{2}}{2m_b}+\mu_{0}\right]b_{\mathbf{q}}+\frac{\lambda^{2}}{2N}\sum_{\mathbf{q}}\mathcal{Q}_{\text{\textbf{q }}}\mathcal{Q}_{-\mathbf{q}},\label{eq:1}
\end{align}
where $\mu_{0}$ and $\lambda$ denote, respectively, the bare bosonic
mass term and the boson-boson interaction, and $\mathcal{Q}_{\mathbf{q}}=\sum_{\mathbf{k}}b_{\mathbf{k}+\mathbf{q}}^{\dagger}b_{\mathbf{k}}$,
where $b_{\mathbf{k}}^{\dagger}$ ($b_{\mathbf{k}}$) is the creation
(annihilation) operator for a boson with momentum $\mathbf{k}$. {The flavor indices are suppressed to not clutter up the notation}. Although
the spin index is not shown for simplicity, we allow for the possibility
that the bosons have either spin-zero or spin-one. As mentioned above,
the model also possesses a ``background'' of fermions (not shown
in the Hamiltonian of Eq. (\ref{eq:1})) and the corresponding fermion-boson
scattering processes lead to retardation effects, which are taken
into account via the one-loop bosonic self-energy $\Sigma\left(\omega\right)=-i\gamma\omega$,
where the Landau-damping constant $\gamma$ is given by ${\gamma=g_{I}^{2}\mathcal{N}(\epsilon_{F})/(2\pi\sqrt{(2k_{F}Q_{0})^{2}-Q_{0}^{4}})}$,
with $\mathcal{N}(\epsilon_{F})$ being the density of states at the
Fermi energy, $k_{F}$ is the corresponding Fermi momentum, $Q_{0}$
is the finite momentum of the bosons and $g_{I}$ is the fermion-boson
interaction. In a previous work \cite{Banerjee20}, we have demonstrated
using the Kubo formula that this effective model indeed displays a quantum
critical phase with approximately $T$-linear resistivity and shows a Drude form for the optical conductivity.

We now proceed to calculate the transport properties of this effective
model within the memory-matrix (MM) formalism \cite{Forster-HFBSCF(1975),Goetze_Woefle,Rosch_Andrei}
(for more information about the technicalities of this method, see,
e.g., Refs. \cite{Hartnoll-PRB_2013,Patel-PRB,Lucas-PRB,Hartnoll_PRB_2014,Freire-AP_2020,Freire-AP_2017,Freire-EPL,Freire-EPL_2018,Freire-AP_2014,Zaanen-CUP,Sachdev-MIT,freire_2020,ips-hermann2,Freire_Mandal_2022,Berg2019,Berg2022}).
The MM approach emerges as a more suitable framework to describe the
non-Fermi liquid phase exhibited, since $(i)$ it does not rely on the existence of well-defined
quasiparticles at low energies, and $(ii)$ it successfully captures
the hydrodynamic regime that is expected to describe the non-equilibrium
dynamics of this strongly correlated metallic phase.

Here, we will follow an approach similar to that used in the recent
work by Wang and Berg \cite{Berg2019} to calculate transport properties
in the context of an Ising-nematic quantum critical theory. In this
way, we will project the non-equilibrium dynamics of the present model
in terms of slowly-varying operators that are nearly conserved. Naturally,
the boson operators $n_{\mathbf{k}}=b_{\mathbf{k}}^{\dagger}b_{\mathbf{k}}$
turn out to be nearly conserved in the limit of either small $\lambda$
or large $N$, since the equation of motion for these operators is
given by 
\begin{align}
\dot{n}_{\mathbf{k}} & =i\left[\hat{{\cal H}},n_{\mathbf{k}}\right]\nonumber \\
 & =\frac{2\lambda^{2}i}{N}\sum_{\mathbf{q}}\mathcal{Q}_{-\mathbf{q}}\left(b_{\mathbf{k}+\mathbf{q}}^{\dagger}b_{\mathbf{k}}-b_{\mathbf{k}}^{\dagger}b_{\mathbf{k}-\mathbf{q}}\right)-h.c..
\end{align}
In the MM formalism, to leading order in $1/N$, the memory matrix
writes 
\begin{align}
M_{\mathbf{k}\mathbf{k}^{\prime}}\left(\Omega\right) & =\frac{1}{i\Omega}\left[G_{\dot{n}_{\mathbf{k}}\dot{n}_{\mathbf{k^{\prime}}}}^{R}\left(\Omega\right)-G_{\dot{n}_{\mathbf{k}}\dot{n}_{\mathbf{k^{\prime}}}}^{R}\left(0\right)\right],\label{eq:3}
\end{align}
where $G_{AB}^{R}\left(\Omega\right)$ is the retarded Green's function
for nearly-conserved operators $A$ and $B$, which is calculated
to zeroth order in $1/N$. The MM turns out to be a generalization of the quasiparticle
scattering rate in Boltzmann theory ({but applicable also to non-Fermi liquids in which this latter quantity cannot be defined}) and enters as
a retardation process in the calculation of the optical conductivity
$\sigma(\Omega)$ and the thermal conductivity at zero electric field
$\bar{\kappa}(\Omega)$ in the following way 
\begin{align}
\sigma\left(\Omega\right) & =\sum_{\mathbf{k}\mathbf{k^{\prime}}}\chi_{J_{\mathbf{k}}n_{\mathbf{k}}}\left(\frac{1}{M_{\mathbf{k}\mathbf{k^{\prime}}}\left(\Omega\right)-i\Omega\chi_{\mathbf{k}\mathbf{k^{\prime}}}}\right)\chi_{J_{\mathbf{k^{\prime}}}n_{\mathbf{k'}}},\label{eq:4}\\
\bar{\kappa}\left(\Omega\right) & =\frac{1}{T}\sum_{\mathbf{k}\mathbf{k^{\prime}}}\chi_{J_{\mathbf{k}}^{\mathcal{Q}}n_{\mathbf{k}}}\left(\frac{1}{M_{\mathbf{k}\mathbf{k^{\prime}}}\left(\Omega\right)-i\Omega\chi_{\mathbf{k}\mathbf{k^{\prime}}}}\right)\chi_{J_{\mathbf{k^{\prime}}}^{\mathcal{Q}}n_{\mathbf{k'}}},\label{eq:5}
\end{align}
where $J_{\mathbf{k}}$ and $J_{\mathbf{k}}^{\mathcal{Q}}$ are, respectively,
the electric current and the thermal current operators of the model,
with the corresponding susceptibilities given by $\chi_{J_{\mathbf{k}}n_{\mathbf{k}}}=\int_{0}^{\beta}d\tau\left\langle J_{\mathbf{k}}\left(\tau\right)n_{\mathbf{k}}\left(0\right)\right\rangle $,
$\chi_{\mathbf{k}\mathbf{k^{\prime}}}=\int_{0}^{\beta}d\tau\left[\left\langle n_{\mathbf{k}}\left(\tau \right)n_{\mathbf{k^{\prime}}}\text{\ensuremath{\left(0\right)}}\right\rangle -\left\langle n_{\mathbf{k}}\right\rangle \left\langle n_{\mathbf{k^{\prime}}}\right\rangle \right]$,
and $\chi_{J_{\mathbf{k}}^{\mathcal{Q}}n_{\mathbf{k}}}=\int_{0}^{\beta}d\tau\left\langle J_{\mathbf{k}}^{\mathcal{Q}}\left(\tau\right)n_{\mathbf{k}}\left(0\right)\right\rangle $.

We point out that the thermal conductivity at zero electric current
of the model (which will be denoted here by $\kappa$) is given by
$\kappa=\bar{\kappa}-T\alpha^{2}/\sigma$, where $\alpha$ is the
thermoelectric coefficient. Since we have demonstrated in a previous work \cite{Banerjee20} that
the present model has particle-hole symmetry, the critical contribution to the thermoelectric
response is expected to vanish. Therefore, in
this case, the thermal conductivity at zero electric current will be equal to the thermal conductivity at zero electric field
(i.e., $\kappa=\bar{\kappa}$).

\begin{figure}[h]
\centering\includegraphics[scale=0.96]{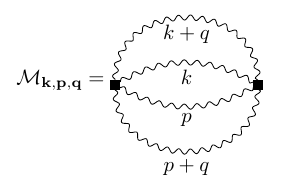} 
\caption{{The Feynman diagram for the calculation of the memory matrix ${\cal M}_{\mathbf{k},\mathbf{p},\mathbf{q}}$ showing up in Eq. (\ref{eq:5.1}) of the present model. The wavy lines are the boson propagator defined in Eq. (\ref{eq:6}). The summation is over the momenta $\mathbf{p}$ and $\mathbf{q}$.}}
\label{fig:Feynman_diagram}
\end{figure}

Furthermore, for clean systems, if no coupling to the lattice is present,
the memory matrix of the model also vanishes identically (see Appendix
\ref{sec:Diagram-contractions}). However, if Umklapp terms are taken
into account, we get after contracting the vertices the following
result:
\begin{align}
M_{\mathbf{k}\mathbf{k^{\prime}}}\left(\Omega\right) =8\,\delta_{\mathbf{k}\mathbf{k^{\prime}}}\sum_{\mathbf{p},\mathbf{q}}{\cal M}_{\mathbf{k},\mathbf{p},\mathbf{q}},\label{eq:5.1}
\end{align}
with the corresponding Feynman diagram given by Fig. \ref{fig:Feynman_diagram}.

\section{Results}

\subsection{Evaluations}

We now compute the various terms in Eqs. (\ref{eq:4}) and (\ref{eq:5}) within our MM
formalism. For convenience, we use units such that $e=k_B=\hbar=1$ from now on. We start from the renormalized propagator for the incoherent bosons given by 
\begin{align}
D_{\mathbf{k}}^{-1} & =\left|\omega_{n}\right|+\mu_{k}(T),\label{eq:6}
\end{align}
with $\mu_{k}(T)=k^{2}+\frac{\lambda^2}{2N}T$. The damping term $\left|\omega_{n}\right|$ (we have set $\gamma=1$)
comes from the scattering via fermionic carriers, and it is responsible
for the incoherent character of the bosons. The potential $\mu_{k}$
has a term $k^{2}$, which refers to the bosonic dispersion. Note that the bosons have a mass scaling with temperature
(the $T$ term in $\mu_{k}$). It comes mainly from the Hartree diagram
generated by the four-boson interaction.

We begin with the evaluation of 
\begin{align}
\chi_{J_{\mathbf{k}}n_{\mathbf{k}}} ={v}_{\mathbf{k}}\int_{0}^{\beta}d\tau\left\langle n_{\mathbf{k}}\left(\tau\right)\right\rangle \left\langle n_{\mathbf{k}}\left(0\right)\right\rangle +{v}_{\mathbf{k}}\chi_{\mathbf{k}\mathbf{k}},\label{eq:7}
\end{align}
with $v_{\mathbf{k}}=\mathbf{k}/m_b$. Henceforth, we define
\begin{align}
&\chi_{J_{\mathbf{k}}n_{\mathbf{k}}}^{(a)}={v}_{\mathbf{k}}\int_{0}^{\beta}d\tau\left\langle n_{\mathbf{k}}\left(\tau\right)\right\rangle \left\langle n_{\mathbf{k}}\left(0\right)\right\rangle,\\
&\chi_{J_{\mathbf{k}}n_{\mathbf{k}}}^{(b)}={v}_{\mathbf{k}}\chi_{\mathbf{k}\mathbf{k}}.
\end{align}
The first term in Eq. (\ref{eq:7}) can be found to be equal to (see
Appendix \ref{subsec:Susceptibilities}) 
\begin{align}
\chi_{J_{\mathbf{k}}n_{\mathbf{k}}}^{(a)} & =\frac{{v}_{\mathbf{k}}n_{B}\left(\mu_{k}\right)}{\pi}\ln\left(\frac{D}{\mu_{k}}\right),\label{eq:7a}
\end{align}
where $D$ is the bandwidth of the boson dispersion and $n_B$ is the Bose-Einstein distribution. Regarding the {second term in Eq. (\ref{eq:7}), we use the 
generalized susceptibility} $\chi_{\mathbf{k}\mathbf{k^{\prime}}}=\delta_{\mathbf{k}\mathbf{k^{\prime}}}T\sum_{\omega_{n}}D_{\mathbf{k}}^{2}$,
which gives (see Appendix \ref{subsec:Susceptibilities}) 
\begin{align}
\chi_{\mathbf{k}\mathbf{k^{\prime}}} & =\delta_{\mathbf{k}\mathbf{k^{\prime}}}\frac{T}{\mu_{k}^{2}}\left(\tanh^{-1}\frac{T}{\mu_{k}}\right)^{2}.\label{eq:8}
\end{align} {The analytical formula of Eq. (\ref{eq:8}) has been obtained in the critical regime where  $\mu_k\lesssim T$. We have used the approximation
$n_{B}\left(x\right)\simeq T/x$ for $\left|x\right|\leq T$, which
is valid for $\mu_{k}\lesssim T$.}

\subsection{Conductivity in the critical regime}

In order to complete the evaluation of the optical conductivity, we
first notice that the summations over $\mathbf{k}$ and $\mathbf{k}^{\prime}$
in Eqs. (\ref{eq:4}) and (\ref{eq:5})  vanish identically if {Umklapp scattering}
is not taken into account {(see Appendix \ref{sec:Diagram-contractions})}. Umklapp terms with $\mathbf{k}^{\prime}=\mathbf{k}\pm n\mathbf{U}$,
with $\mathbf{U}$ being a reciprocal lattice wave vector, generate
a finite result for the optical conductivity. This result is obtained
by the scaling displayed in Eqs. (\ref{eq:7})-(\ref{eq:9}) in the
critical regime. In this regime, the second term $\chi_{J_{\mathbf{k}}n_{\mathbf{k}}}^{(b)}$
in Eq. (\ref{eq:7}) dominates over the first term (see
Appendix \ref{subsec:Remarks}). Moreover, the
MM evaluates to 
\begin{align}
M_{\mathbf{k}\mathbf{k^{\prime}}}\left(\Omega_{n}\right)=\delta_{\mathbf{k}\mathbf{k^{\prime}}}\frac{\lambda^{4}T^{2}}{N^{2}\Omega_{n}}\text{\ensuremath{\sum_{\omega_{n},p_{n}}\sum_{\mathbf{p},\mathbf{q}}D_{\mathbf{p}}D_{\mathbf{k}}D_{\mathbf{p}+\mathbf{q}}D_{\mathbf{k+}\mathbf{q}}}},
\end{align}
which finally yields (see
Appendix \ref{subsec:Memory})
\begin{align}
M_{\mathbf{k}\mathbf{k^{\prime}}}\left(\Omega\right) & =\frac{\delta_{\mathbf{k}\mathbf{k}^{\prime}}\lambda^{4}}{N^{2}}\frac{T^{3}}{192\pi^{3}\mu_{k}^{3}}\left(\frac{\mu_{k}T}{\mu_{k}^{2}+T^{2}}+\tanh^{-1}\frac{T}{\mu_{k}}\right).\label{eq:9}
\end{align}
Lastly, the optical conductivity can be rewritten as 
\begin{align}
\sigma\left(\Omega\right) & =\sum_{\mathbf{k}}\frac{\chi_{\mathbf{k}\mathbf{k}}}{\chi_{\mathbf{k}\mathbf{k}}^{-1}M_{\mathbf{k}\mathbf{k}}\left(\Omega\right)-i\Omega}.\label{eq:10}
\end{align}
Noticing that the typical scaling relation $\mu_{k}\sim T$  holds
in the critical regime (because $T$ is the only energy scale in the problem and thus $k^2\sim T$), the summation over $\mathbf{k}$ in Eq. (\ref{eq:10})
can be finally performed. Scalings arguments lead to $\sum_{\mathbf{k}}\sim T$,
$M_{\mathbf{k}\mathbf{k}}\sim T^{0}$, $\chi_{\mathbf{k}\mathbf{k}}\sim T^{-1}$,
which result in a typical form for the optical conductivity given
by 
\begin{align}
\sigma\left(\Omega\right) & \sim\frac{1}{T-i\Omega}.\label{eq:11}
\end{align}
The aforementioned result has been further validated in the dc limit by numerically summing over $\mathbf{k}$ in Eq. \eqref{eq:10}, which confirms our results in Ref. \cite{Banerjee20}.

\subsection{Lorenz ratio in the critical regime}

The Wiedemann-Franz law \cite{chester1961law} for the Lorenz ratio $\text{L}=\frac{\kappa}{\sigma T}$
is one of the most fundamental properties of a Fermi liquid. It states
that at low temperatures 
\begin{align}
\lim_{T\rightarrow0}\text{L}=\frac{\pi^{2}}{3}\equiv\text{L}_{0},\label{eq:12}
\end{align}
in units with $k_{B}=e=1$. It reflects the fact that energy and charge
are carried by the same degrees of freedom. To compute this ratio, we compute
the thermal conductivity $\kappa$ using the same method with the following substitution of the susceptibility $\chi_{J_{\mathbf{k}}^{\mathcal{Q}}n_{{\mathbf{k}}}}=\epsilon_{{\mathbf{k}}}\chi_{J_{\mathbf{k}}n_{{\mathbf{k}}}}$.
At the critical regime, we get within the MM approach 
\begin{align}
\kappa\left(\Omega\right) & =\frac{1}{T}\sum_{\mathbf{k}}\epsilon_\mathbf{k}^2\frac{\chi_{\mathbf{k}\mathbf{k}}}{\chi_{\mathbf{k}\mathbf{k}}^{-1}M_{\mathbf{k}\mathbf{k}}\left(\Omega\right)-i\Omega}.\label{eq:13}
\end{align}
Scaling arguments lead to $\epsilon_{\mathbf{k}}\sim T$, which finally
gives for the thermal conductivity 
\begin{align}
\kappa\left(\Omega\right) & \sim\frac{T}{T-i\Omega}.\label{eq:14}
\end{align}
In the critical regime, the incoherent boson system obeys the correct scaling, $\frac{\kappa}{\sigma T}\sim C$, with $C$ being a constant
as $T\rightarrow0$. {However,  although the Lorenz ratio is a constant, it does not satisfy the WF law because the coefficient is strongly dependent on the boson-boson coupling $\dfrac{\lambda^2}{2 N}$ (as shown in Fig. \ref{fig:Lorentz}) and so is model-dependent. This result is also confirmed using Kubo linear response in Appendix \ref{sec:Kubo}. In the $t-J$ model \cite{houghton2002violation} and near heavy fermion quantum critical point \cite{kim2009violation}, stronger violations have been observed, where the Lorenz ratio does not saturate to a constant at low temperatures.}\\

{In the cuprates, the experimental situation is similar to our findings. The Lorenz ratio is found to saturate to a constant, both in the overdoped \cite{proust2002heat,nakamae2003electronic,Cyril02_PRL,WFPRX18} and underdoped \cite{grissonnanche2016wiedemann} regimes, but the value of this constant depends on the oxygen doping. This experimental fact thus constrains the bosonic coupling $\lambda$ of our model. } We point out that this result can also be traced to the fact that we considered Umklapp scattering as the sole mechanism for momentum relaxation in the present calculation.

\begin{figure}[t]
\includegraphics[scale=0.55]{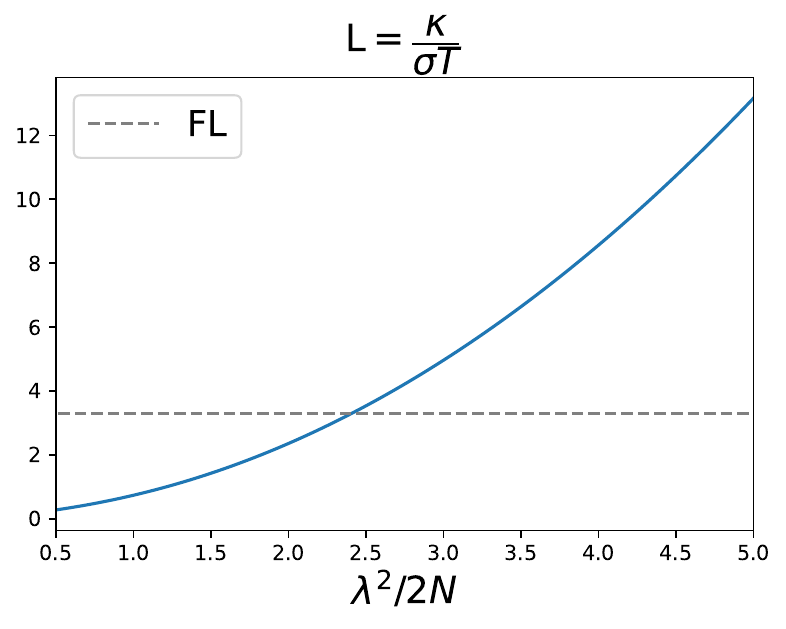} \caption{The Lorenz ratio $\text{L}=\kappa/(\sigma T)$ (in units with $k_{B}=e=1$) as a function of the bosonic coupling $g=\dfrac{\lambda^2}{2N}$, computed
within the Kubo formalism. The same results are obtained with the memory matrix formalism. $T$ is fixed such that it satisfies $T>\dfrac{\mu_0}{g}$, in order to be in the critical regime for all coupling values used.}
\label{fig:Lorentz} 
\end{figure}

\section{Conclusions}

In this paper, we have computed the transport properties of an effective model
when charged fluctuating bosons (charge-two particle-particle bosons
in this special case) are present at high energies in the phase diagram
of the cuprate superconductors. The main results are as follows:
\begin{itemize}
\item A regime with approximately $T$-linear resistivity is obtained from the
transport properties of the boson-fermion ``soup''. The charged
bosons scatter with the fermions and become overdamped via the Landau
damping $-i\gamma\omega$ (where $\omega$ is the real frequency).
Optical conductivity was also evaluated \cite{Banerjee20} and yields a Drude-like conductivity for $\omega< T$
with a lifetime given by $\tau_{xx}=\tau_{b}\sim T^{-1}$. This agrees with experimental observation \cite{michon2023reconciling}.

\item At low temperatures, the boson transport in the conductivity is ``short-circuited''
by the fermions that possess a transport lifetime $\tau_{f}\sim T^{-2}$. Hence, the
regime where the bosons dominate the conductivity has a finite temperature
range, which needs to be compared with the experimental data (this will be performed below).

\item Due to  {the particle-hole symmetry of the Landau damped bosons}, those charged bosons do not contribute to
the Hall conductivity. {This finding was already part of our previous study~\cite{Banerjee20}  and is also in good agreement with the experimental study of Ref.~\cite{putzke2019reduced}. Therefore, the fermions dominate the Hall conductivity via
hot-spot and cold-spot physics. A similar situation was studied in \cite{Hussey_McKenzie}, where a linear-in-$T$ longitudinal resistivity was assumed in parallel with ``hot spot" and ``cold spot" physics to estimate the Hall conductivity. Remarkably, their phenomenological analysis agreed with the experimental data and showed that the averaging around the Fermi surface for the Hall conductivity integral was weighting the hot spots more than for the longitudinal conductivity. A simplified understanding of their results in terms of a ``lifetime picture"  would give for the Hall conductivity from the
fermions} described by $\tau_{xy}\sim T^{-3/2}$, since the Hall average around the Fermi surface scans at the same time both the hot and
cold regions. Altogether, in this regime, the
{cotangent  of the Hall  angle } goes as $\cot\theta_H\sim \tau_{xx}/\tau_{xy}^2\sim T^{3}/T\sim T^{2}$, which
corresponds to the experimental observation.

\item The thermal transport has also been calculated, and in the regime analyzed here (i.e., the critical case), the {strict Wiedemann-Franz law (with the universal coefficient from the Fermi liquid theory) }is violated at low temperatures due to the dependence on the bosonic coupling. In other words, although the Lorenz ratio does converge to a constant at low temperatures, the WF law is violated in view of its dependence on the bosonic coupling. By constraining this coupling, the WF law can {be brought to agree }with recent experimental data \cite{WFPRX18}, but as mentioned in the main text, there is still room for improvement here. It would be interesting to investigate also the effects of adding disorder via spatially random interactions in our model.
\end{itemize}

Finally, we point out that although our present study might not yet be the final solution
for the strange metal phase of the cuprates, this perspective
opens a new viewpoint on the physics of those compounds. The main
idea is that at high energy scales, due to the strong superexchange
interaction which brings the system to the regime of strong coupling,
bosons of charge-zero (particle-hole) and charge-two (particle-particle)
with a spectrum of wavevectors are generated. When the temperature
is lowered, some of those bosons condense at a specific wave vector,
giving rise to various orders, like, e.g., charge modulations
or stripe physics depending on the compounds. Some bosons are unstable,
like the charge-two finite vector ones related, e.g., to pair-density wave
(PDW) order. In a recent line of ideas \cite{WFPRB16, Grandadam19, Banerjee21}, the instability of the finite momentum
charge-two bosons could be described by a theory of ``fractionalization.''
in which these excitations become entangled, opening a gap in the antinodal region
of the Brillouin zone. In the strange metal phase, they also give rise
to the regime of $T$-linear resistivity described in the present paper.

As mentioned above, it is worth plugging numbers to determine the possible boundaries of the obtained strange metal regime, which is valid above $T>T_{min}$. We point out that the constraint on $T_{min}$ is such that the fermions should
short-circuit the bosons for $T<T_{min}$. So we may impose that, for $T\lesssim T_{min}$, we have that
$\sigma_{f}\gtrsim\sigma_{b}$. For a rough estimate, we can then use for the fermions that $\sigma_{f}=ne^{2}\tau_{f}/m_ f^{*}$,
where $m_f^{*}$ is the quasiparticle effective mass. For the bosons, since we showed{ by means of scaling arguments within the MM formalism here and also using Kubo formalism in Ref. \cite{Banerjee20} } that they also obey the Drude form for the conductivity, we have that
$\sigma_{b}=n_{b}\left(2e\right)^{2}\tau_{b}/m_b$, where $m_b$ is the
mass introduced in Eq. (\ref{eq:1}). From {Fig. 2 of }Ref. \cite{Hussey_McKenzie}, 
one obtains that $\sigma_{f}^{-1}=A\,T^{2}$, with $A\simeq5\cdot 10^{-3}\mu\Omega\cdot \text{cm}$.
Thus, the reasonable range allowed experimentally for $T_{min}$ is $1 \text{ K}<T_{min}<10 \text{ K}$.
If $T_{min}=1 \text{ K}$, one would get $m_b/m_f^{*}\sim0.02$, whereas if $T_{min}=10 \text{ K}$
the ratio of masses should be given by $m_b/m_f^{*}\sim0.2$. Given the
strong mass renormalization of the fermions inside the strange metal phase of the present model, this range of the ratio of masses is conceivable 
in order to allow for a wide fluctuation regime where the incoherent bosons dominate the transport properties with respect to
the fermions in the context of the resistivity of this non-Fermi liquid phase.

\section*{Acknowledgments}

We are grateful for discussion with G. Grissonnanche, N. Hussey, B.
Ramshaw, L. Taillefer on various experimental issues. H.F. acknowledges
funding from CNPq under Grant No. 311428/2021-5. A.B. acknowledges support from the Kreitman School of Advanced Graduate Studies and European Research Council (ERC) Grant Agreement No. 951541, ARO (W911NF-20-1-0013).

\appendix

\section{Diagram contractions\label{sec:Diagram-contractions}}

We define the vertices as follows 
\begin{figure}[h]
\centering\includegraphics[scale=0.85]{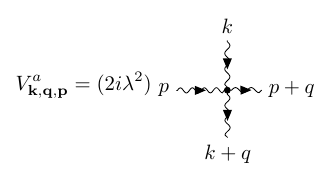} 
\centering\includegraphics[scale=0.85]{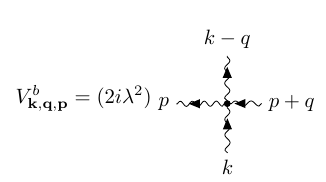} 
\end{figure}

\noindent and their conjugate counterpart. The vertices write 
\begin{align}
 & V_{\mathbf{k}}=\sum_{\mathbf{p},\mathbf{q}}V_{\mathbf{k},\mathbf{p},\mathbf{q}},\label{eq:A1}\\
 & \mbox{with }V_{\mathbf{k,}\mathbf{p},\mathbf{q}}=V_{\mathbf{k},\mathbf{p},\mathbf{q}}^{a}-V_{\mathbf{k},\mathbf{p},\mathbf{q}}^{b}-h.c..\nonumber 
\end{align}
We construct the function ${\cal M}_{\mathbf{k},\mathbf{p},\mathbf{q}}$
by contracting 
\begin{align}
{\cal M}_{\mathbf{k},\mathbf{p},\mathbf{q}} & =\left\langle V_{\mathbf{k},\mathbf{p},\mathbf{q}}\cdot\bar{V}_{\mathbf{k^{\prime},}\mathbf{p^{\prime}},\mathbf{q^{\prime}}}\right\rangle ,\label{eq:A2}\\
 & =\left<-V_{\mathbf{k},\mathbf{p},\mathbf{q}}^{a}\cdot V_{\mathbf{k^{\prime}},\mathbf{p^{\prime}},\mathbf{q^{\prime}}}^{a}-V_{\mathbf{k},\mathbf{p},\mathbf{q}}^{b}\cdot V_{\mathbf{k^{\prime}},\mathbf{p^{\prime}},\mathbf{q^{\prime}}}^{b}\right.\nonumber \\
 & +\bar{V}_{\mathbf{k},\mathbf{p},\mathbf{q}}^{a}\cdot V_{\mathbf{k^{\prime}},\mathbf{p^{\prime}},\mathbf{q^{\prime}}}^{a}+\bar{V}_{\mathbf{k},\mathbf{p},\mathbf{q}}^{b}\cdot V_{\mathbf{k^{\prime}},\mathbf{p^{\prime}},\mathbf{q^{\prime}}}^{b}\nonumber \\
 & +V_{\mathbf{k},\mathbf{p},\mathbf{q}}^{a}\cdot V_{\mathbf{k^{\prime}},\mathbf{p^{\prime}},\mathbf{q^{\prime}}}^{b}+V_{\mathbf{k},\mathbf{p},\mathbf{q}}^{b}\cdot V_{\mathbf{k^{\prime}},\mathbf{p^{\prime}},\mathbf{q^{\prime}}}^{a}\nonumber \\
 & -\left.\bar{V}_{\mathbf{k},\mathbf{p},\mathbf{q}}^{a}\cdot V_{\mathbf{k^{\prime}},\mathbf{p^{\prime}},\mathbf{q^{\prime}}}^{b}-\bar{V}_{\mathbf{k},\mathbf{p},\mathbf{q}}^{b}\cdot V_{\mathbf{k^{\prime}},\mathbf{p^{\prime}},\mathbf{q^{\prime}}}^{a}+h.c.\right>.\nonumber 
\end{align}
There are eight types of contractions 
\begin{align}
1) & \left\langle V_{\mathbf{k},\mathbf{p},\mathbf{q}}^{a}\cdot V_{\mathbf{k^{\prime}},\mathbf{p^{\prime}},\mathbf{q^{\prime}}}^{a}\right\rangle =\delta_{\mathbf{k},\mathbf{k^{\prime}}-\mathbf{q}}\delta_{\mathbf{p^{\prime},}\mathbf{p}+\mathbf{q}}\delta_{\mathbf{q^{\prime},-}\mathbf{q}}{\cal M}_{\mathbf{k},\mathbf{p},\mathbf{q}},\label{eq:A3}\\
2) & \left\langle V_{\mathbf{k},\mathbf{p},\mathbf{q}}^{b}\cdot V_{\mathbf{k^{\prime}},\mathbf{p^{\prime}},\mathbf{q^{\prime}}}^{b}\right\rangle =\delta_{\mathbf{k},\mathbf{k^{\prime}}+\mathbf{q}}\delta_{\mathbf{p^{\prime},}\mathbf{p}+\mathbf{q}}\delta_{\mathbf{q^{\prime},-}\mathbf{q}}{\cal M}_{\mathbf{k},\mathbf{p},\mathbf{q}},\nonumber \\
3) & \left\langle V_{\mathbf{k},\mathbf{p},\mathbf{q}}^{a}\cdot V_{\mathbf{k^{\prime}},\mathbf{p^{\prime}},\mathbf{q^{\prime}}}^{b}\right\rangle =\delta_{\mathbf{k},\mathbf{k^{\prime}}}\delta_{\mathbf{p^{\prime},}\mathbf{p}+\mathbf{q}}\delta_{\mathbf{q^{\prime},-}\mathbf{q}}{\cal M}_{\mathbf{k},\mathbf{p},\mathbf{q}},\nonumber \\
4) & \left\langle V_{\mathbf{k},\mathbf{p},\mathbf{q}}^{b}\cdot V_{\mathbf{k^{\prime}},\mathbf{p^{\prime}},\mathbf{q^{\prime}}}^{a}\right\rangle =\left\langle V_{\mathbf{k},\mathbf{p},\mathbf{q}}^{a}\cdot V_{\mathbf{k^{\prime}},\mathbf{p^{\prime}},\mathbf{q^{\prime}}}^{b}\right\rangle ,\nonumber \\
5) & \left\langle \bar{V}_{\mathbf{k},\mathbf{p},\mathbf{q}}^{a}\cdot V_{\mathbf{k^{\prime}},\mathbf{p^{\prime}},\mathbf{q^{\prime}}}^{a}\right\rangle =\delta_{\mathbf{k},\mathbf{k^{\prime}}}\delta_{\mathbf{p^{\prime},}\mathbf{p}}\delta_{\mathbf{q^{\prime},q}}{\cal M}_{\mathbf{k},\mathbf{p},\mathbf{q}},\nonumber \\
6) & \left\langle \bar{V}_{\mathbf{k},\mathbf{p},\mathbf{q}}^{b}\cdot V_{\mathbf{k^{\prime}},\mathbf{p^{\prime}},\mathbf{q^{\prime}}}^{b}\right\rangle =\left\langle \bar{V}_{\mathbf{k},\mathbf{p},\mathbf{q}}^{a}\cdot V_{\mathbf{k^{\prime}},\mathbf{p^{\prime}},\mathbf{q^{\prime}}}^{a}\right\rangle ,\nonumber \\
7) & \left\langle \bar{V}_{\mathbf{k},\mathbf{p},\mathbf{q}}^{a}\cdot V_{\mathbf{k^{\prime}},\mathbf{p^{\prime}},\mathbf{q^{\prime}}}^{b}\right\rangle =\delta_{\mathbf{k},\mathbf{k^{\prime}}-\mathbf{q}}\delta_{\mathbf{p^{\prime},}\mathbf{p}+\mathbf{q}}\delta_{\mathbf{q^{\prime},q}}{\cal M}_{\mathbf{k},\mathbf{p},\mathbf{q}},\nonumber \\
8) & \left\langle \bar{V}_{\mathbf{k},\mathbf{p},\mathbf{q}}^{b}\cdot V_{\mathbf{k^{\prime}},\mathbf{p^{\prime}},\mathbf{q^{\prime}}}^{a}\right\rangle =\delta_{\mathbf{k},\mathbf{k^{\prime}}+\mathbf{q}}\delta_{\mathbf{p^{\prime},}\mathbf{p}+\mathbf{q}}\delta_{\mathbf{q^{\prime},q}}{\cal M}_{\mathbf{k},\mathbf{p},\mathbf{q}}.\nonumber 
\end{align}
Altogether, we finally have 
\begin{align}
M_{\mathbf{k}\mathbf{k^{\prime}}}\left(\Omega\right) & =4\sum_{\mathbf{p},\mathbf{q}}\left(\delta_{\mathbf{k},\mathbf{k^{\prime}}}-\delta_{\mathbf{k},\mathbf{k^{\prime}}\pm\mathbf{q}}\right){\cal M}_{\mathbf{k},\mathbf{p},\mathbf{q}},\label{eq:A4}
\end{align}
which vanishes if {Umklapp scattering is not taken into account}. Umklapp terms
provide a nonzero value to $M_{\mathbf{k}\mathbf{k^{\prime}}}\left(\Omega\right)$,
leading to Eq. (\ref{eq:9}) presented in the main part of the manuscript.

\section{Evaluations\label{sec:Evaluations}}

\subsection{Susceptibilities\label{subsec:Susceptibilities}}

After performing a spectral decomposition of the boson Green's function
Eq. (\ref{eq:6}), the first part of the susceptibility $\chi_{J_{\mathbf{k}}n_{\mathbf{k}}}$
writes 
\begin{align}
\chi_{J_{\mathbf{k}}n_{\mathbf{k}}}^{(a)} & =\mathbf{v}_{k}\int_{-D}^{D}\frac{dE}{2\pi}n_{B}\left(E\right)\frac{E}{E^{2}+\mu_{k}^{2}}\label{eq:B1}\\
 & =\mathbf{v}_{k}\frac{n_{B}\left(\mu_{k}\right)}{\pi}\ln\left(\frac{D}{\mu_{k}}\right),\nonumber 
\end{align}
with the convention {$n_{B}\left(E\right)=\exp{\delta E}/\left ( \exp{\beta E}-1\right )$, with $\delta\ll 1$ being a regulator for negative energies.}

As for the susceptibility $\chi_{\mathbf{k}\mathbf{k}}$, we proceed
in the same way with a spectral decomposition of each boson propagator
given by 
\begin{align}
\chi_{\mathbf{k}\mathbf{k}} & =\int\frac{dE_{1}dE_{2}}{\left(2\pi\right)^{2}}\frac{n_{B}\left(E_{1}\right)-n_{B}\left(E_{2}\right)}{-E_{1}+E_{2}}\frac{E_{1}}{E_{1}^{2}+\mu_{k}^{2}}\frac{E_{2}}{E_{2}^{2}+\mu_{k}^{2}}\label{eq:B2}\\
 & \simeq\int_{0}^{T}\frac{dE_{1}dE_{2}}{\left(2\pi\right)^{2}}\frac{T}{\left(E_{1}^{2}+\mu_{k}^{2}\right)\left(E_{2}^{2}+\mu_{k}^{2}\right)}\nonumber \\
 & =\frac{T}{\mu_{k}^{2}}\left(\tanh^{-1}\frac{T}{\mu_{k}}\right)^{2},\nonumber 
\end{align}
where in the {second line we have used the approximation
$n_{B}\left(x\right)\simeq T/x$ for $\left|x\right|\leq T$, which
is valid for $\mu_{k}\lesssim T$.}

\subsection{Remarks on the scaling of the optical conductivity\label{subsec:Remarks}}

{In the scaling of the optical conductivity of Eq. (\ref{eq:4}),
the two terms $\chi_{J_{\mathbf{k}}n_{\mathbf{k}}}^{(a)}$ and $\chi_{J_{\mathbf{k}}n_{\mathbf{k}}}^{(b)}$
behave differently. Any term proportional to $\chi_{J_{\mathbf{k}}n_{\mathbf{k}}}^{(a)}$
in the summation over the momentum in Eq. (\ref{eq:4}) gives a UV
divergence (after turning the summation over the momentum into an integral in the thermodynamic limit). In other words, the typical momentum
of the integral is such that $\mu_{k}\sim D$, where $D$ is the bandwidth.
Then, the factor $n_{B}\left(\mu_{k}\right)\sim n_{B}\left(D\right)$
in Eq. (\ref{eq:7}) leads to an exponentially small contribution.
On the other hand, the contribution related to $\chi_{J_{\mathbf{k}}n_{\mathbf{k}}}^{(b)}$
in the optical conductivity integral in Eq. (\ref{eq:4}) is dominated
by a typical momentum such that $\mu_{k}\sim T$. This leads to the
result of Eq. (\ref{eq:11}).}

\subsection{Memory Matrix\label{subsec:Memory}}

The MM, after spectral decomposition, writes 
\begin{align}
M_{\mathbf{k}\mathbf{k}}\left(\Omega_{n}\right) & =\frac{\lambda^{4}}{N^{2}}\frac{1}{\Omega_{n}}T\sum_{\omega_{q}}\sum_{\mathbf{p},\mathbf{q}}\int\prod_{i=1}^{4}\frac{dE_{i}}{2\pi}F\left(E_{i}\right)\nonumber \\
 & \times\frac{n_{B}\left(E_{3}\right)-n_{B}\left(E_{4}\right)}{E_{3}-E_{4}+i\omega_{q}}\frac{n_{B}\left(E_{1}\right)-n_{B}\left(E_{2}\right)}{E_{1}-E_{2}+i\omega_{q}+i\Omega_{n}},\nonumber \\
\end{align}
with $F\left(E\right)=\frac{E}{E^{2}+\mu_{k}^{2}}$. Performing the
sum over $\omega_{q}$, we get 
\begin{align}
M_{\mathbf{k}\mathbf{k}}\left(\Omega_{n}\right) & =\frac{\lambda^{4}}{N^{2}}\frac{1}{\Omega_{n}}\sum_{\mathbf{p},\mathbf{q}}\int\prod_{i=1}^{4}\frac{dE_{i}}{2\pi}F\left(E_{i}\right)\nonumber \\
 & \times\left[n_{B}\left(E_{1}\right)-n_{B}\left(E_{2}\right)\right]\left[n_{B}\left(E_{3}\right)-n_{B}\left(E_{4}\right)\right]\nonumber \\
 & \times\frac{n_{B}\left(E_{1}-E_{2}\right)-n_{B}\left(E_{3}-E_{4}\right)}{E_{1}-E_{2}-E_{3}+E_{4}+i\Omega_{n}}.
\end{align}
We now perform analytic continuation $i\Omega_{n}\rightarrow\Omega+i$$\delta$.
We obtain, after taking the limit $\Omega\rightarrow0$, 
\begin{align}
M_{\mathbf{k}\mathbf{k}}\left(\Omega\right) & =\frac{\lambda^{4}}{N^{2}}\pi\sum_{\mathbf{p},\mathbf{q}}\int\prod_{i=1}^{4}\frac{dE_{i}}{2\pi}F\left(E_{i}\right)\nonumber \\
 & \times\left[n_{B}\left(E_{1}\right)-n_{B}\left(E_{2}\right)\right]\left[n_{B}\left(E_{3}\right)-n_{B}\left(E_{4}\right)\right]\nonumber \\
 & \times\delta\left(E_{1}-E_{2}-E_{3}+E_{4}\right)\frac{\partial n_{B}}{\partial E}\bigg|_{E_{1}-E_{2}}.\nonumber \\ \label{eq:B5}
\end{align}
We now use $\partial n_{B}/\partial E\simeq-T/E^{2}$, if $\left|E\right|<T$
and zero elsewhere. We have a factor $-T/(E_{1}-E_{2})^{2}$, with
the condition that $\left|E_{1}-E_{2}\right|<T$. We thus have $E_{2}\simeq E_{1}\pm T$
and form resolving the $\delta$-function, $E_{4}\simeq E_{3}\pm T$.
This in turn gives $n_{B}\left(E_{1}\right)-n_{B}\left(E_{2}\right)\simeq\left(\partial n_{B}/\partial E_{1}\right)T\simeq-T^{2}/E_{1}^{2}$
and likewise $n_{B}\left(E_{3}\right)-n_{B}\left(E_{4}\right)\simeq-T^{2}/E_{3}^{2}$.
Eliminating two variables in the integral Eq. (\ref{eq:B5}), but
remembering that $\left|E_{1}-E_{2}\right|<T$ and $\left|E_{3}-E_{4}\right|<T$,
leads to

\begin{align}
M_{\mathbf{k}\mathbf{k}}\left(\Omega\right) & =\frac{\lambda^{4}}{N^{2}}\frac{\pi}{\left(2\pi\right)^{4}}\sum_{\mathbf{p},\mathbf{q}}\int_{-T}^{T}\int_{-T}^{T}dE_{1}dE_{3}T^{4}\nonumber \\
 & \times\frac{1}{\left(E_{1}^{2}+\mu_{p}^{2}\right)\left(E_{1}^{2}+\mu_{p+q}^{2}\right)\left(E_{3}^{2}+\mu_{k}^{2}\right)\left(E_{3}^{2}+\mu_{k+q}^{2}\right)}\nonumber \\
 & =\frac{\lambda^{4}}{N^{2}}\frac{\pi}{\left(2\pi\right)^{4}}T^{4}\sum_{\mathbf{p},\mathbf{q}}I_{\mathbf{p},\mathbf{q}}I_{\mathbf{k},\mathbf{q}}.
\end{align}
With $\left(a,b\right)=\left(\mu_{p},\mu_{p+q}\right)$ for the first
integral and $\left(a,b\right)=\left(\mu_{k},\mu_{k+q}\right)$ for
the second integral, we obtain the following result 
\begin{align}
I_{a,b} & =2\int_{0}^{T}\frac{dx}{\left(x^{2}+a^{2}\right)\left(x^{2}+b^{2}\right)}\nonumber \\
 & =2\left[\frac{-b\tan^{-1}\frac{T}{a}+a\tan^{-1}\frac{T}{b}}{ab\left(a^{2}-b^{2}\right)}\right].
\end{align}
In order to go further, we need to perform the integration over $\mathbf{p}$
and $\mathbf{q}$. For this reason, we assume (and this can be also
checked at the end of the calculation) that the order of magnitude
of the various wave vectors is such that they are all scaling with temperature
as $k\sim q\sim p\sim\sqrt{T}$. We can thus use $\mu_{k+q}\simeq\mu_{k}$
inside the integrals. We then have 
\begin{align}
M_{\mathbf{k}\mathbf{k}}\left(\Omega\right) & =\frac{\lambda^{4}}{N^{2}}\frac{\pi}{\left(2\pi\right)^{4}}T^{4}J_{\mathbf{p},\mathbf{q}}\frac{1}{\mu_{k}^{3}}\left[\frac{\mu_{k}T}{\mu_{k}^{2}+T^{2}}+\tan^{-1}\frac{T}{\mu_{k}}\right],\label{eq:B7}
\end{align}
with 
\begin{align}
J_{\mathbf{p},\mathbf{q}} & =\int_{0}^{T}\int_{0}^{T}\frac{dp^{2}dq^{2}}{\left(\mu_{p}^{2}-\mu_{p+q}^{2}\right)}\left[-\frac{1}{\mu_{p+q}}\tan^{-1}\frac{T}{\mu_{p}}\right.\nonumber\\
&\left.+\frac{1}{\mu_{p}}\tan^{-1}\frac{T}{\mu_{p+q}}\right].
\end{align}
Recalling that $\mu_{p}=p^{2}+T$ and $\mu_{p+q}=\left(p+q\right)^{2}+T$,
and expanding $\tan^{-1}\left(T/\mu_{p}\right)\sim T/\mu_{p}-\left(T/\mu_{p}\right)^{3}$,
we get 
\begin{align}
J_{\mathbf{p},\mathbf{q}} & \simeq\frac{T^{3}}{3}\int_{0}^{T}\frac{dp^{2}d(p+q)^{2}}{\mu_{p}^{3}\mu_{p+q}^{3}}\nonumber \\
 & =\frac{1}{12T}.\label{eq:B8}
\end{align}
Putting together Eqs. (\ref{eq:B7}) and (\ref{eq:B8}), it leads
to 
\begin{align}
M_{\mathbf{k}\mathbf{k}}\left(\Omega\right) & =\frac{\lambda^{4}}{N^{2}}\frac{\pi}{12\left(2\pi\right)^{4}}\frac{T^{3}}{\mu_{k}^{3}}\left(\frac{\mu_{k}T}{\mu_{k}^{2}+T^{2}}+\tanh^{-1}\frac{T}{\mu_{k}}\right),\label{eq:B9}
\end{align}
which is the result presented in the main part of the manuscript.

\section{Comparison with Kubo formalism\label{sec:Kubo}}

\begin{figure}[t]
\includegraphics[scale=0.55]{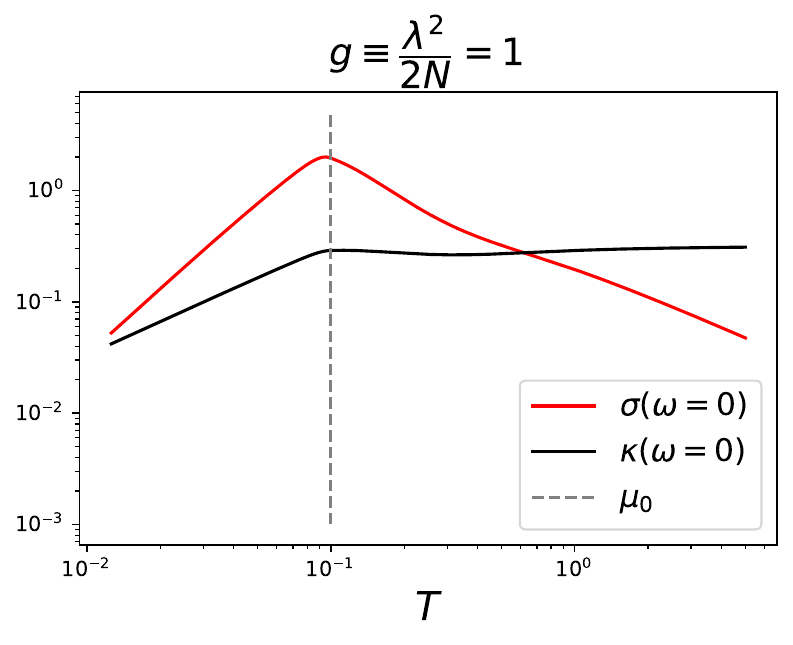}
\caption{The electrical conductivity and thermal conductivity as a function of $T$ computed
within Kubo formalism in units with $e=1$ and $\hbar=1$. Here, we set $g\equiv\frac{\lambda^2}{2N}=1$. For $T>\mu_{0}$, in the critical regime, the constant behavior
of $\kappa$ and the $\dfrac{1}{T}$ behavior of $\sigma$ set in.}
\label{fig:final} 
\end{figure}

Here, we compare the results obtained within the memory matrix formalism
with the standard Kubo formalism \cite{kubo1957statistical}. Following our previous work \cite{Banerjee20},
the optical conductivity and the thermal conductivity (as we pointed
out in the main text, the thermal conductivity at zero electric
field $\bar{\kappa}$ and the thermal conductivity at zero electric current $\kappa$ are equal
in the present calculation) are given, respectively, by 
\begin{align}
\sigma(\omega)&=-\dfrac{\mathcal{K}_{0}(\omega_{n})}{\omega_{n}}\bigg|_{i\omega_{n}\rightarrow\omega+i0^{+}},\\
{\kappa}(\omega)&=-\dfrac{1}{T}\dfrac{\mathcal{K}_{2}(\omega_{n})}{\omega_{n}}\bigg|_{i\omega_{n}\rightarrow\omega+i0^{+}},
\end{align}
with $\mathcal{K}_{\alpha}$ computed using the spectral decomposition of the boson propagator $D^{-1}(\mathbf{q},\omega_n)=|\omega_n|+\mu_\mathbf{q}(T)$ used in Eq. (\ref{eq:B1})
\begin{align}
\nonumber&\mathcal{K}_\alpha(\omega_n)=&\\
&-\sum\limits_{\mathbf{q}}\epsilon_{\mathbf{q}}^{\alpha}\int\frac{dE_{1}dE_{2}}{\left(2\pi\right)^{2}}\frac{n_{B}\left(E_{1}\right)-n_{B}\left(E_{2}\right)}{-E_{1}+E_{2}+\omega_n}\frac{E_{1}}{E_{1}^{2}+\mu_{\mathbf{q}}^{2}}\frac{E_{2}}{E_{2}^{2}+\mu_{\mathbf{q}}^{2}}.&
\end{align}

Performing an analytical continuation $\omega_n \rightarrow \omega+i\delta$ and considering the limit $\omega \rightarrow 0$, we get:
\begin{align}
&\mathcal{K}_\alpha(\omega)=\dfrac{T\omega}{4\pi}\sum\limits_{\mathbf{q}}\epsilon_{\mathbf{q}}^{\alpha}\int\frac{dE}{\left(2\pi\right)^{2}}\frac{T}{E^2}\left(\frac{E}{E^{2}+\mu_{\mathbf{q}}^{2}}\right)^2.&
\end{align}
$\mathcal{K}_{\alpha}$ can be integrated explicitly and we obtain the optical and thermal conductivities $\sigma$ and $\kappa$ shown in Fig. \ref{fig:final}.

The $T$-dependence of $\mu$ is given by
\begin{equation}
\mu\equiv\mu(T)=\left\{ \begin{array}{ll}
\mu_{0}+\dfrac{\lambda^{2}}{2N}T\,\ln\left(\dfrac{T}{\mu_{0}}\right) & \text{for }T\gg\mu_{0},\\
\mu_{0} & \text{for }T\ll\mu_{0}.
\end{array}\right.
\end{equation}
We consider the critical regime to compare our results with those obtained from the memory matrix formalism. $T$ satisfies $T>\mu_0$ and $T>\dfrac{\lambda^2\mu_0}{2N}$; thus, we can approximate $\dfrac{\mu}{T} \approx \dfrac{\lambda^2}{2N}$ by neglecting logarithmic corrections. In this regime, we have $\sigma=\dfrac{1}{T}\tilde{\sigma}(\mu/T)$ and $\kappa=\tilde{\kappa}(\mu/T)$, in agreement with the scaling obtained using the memory matrix formalism in the critical regime in Eqs. (\ref{eq:11}) and (\ref{eq:14}). If we consider logarithmic corrections in $\mu(T)$, the results obtained remain qualitatively similar up to a weak dependence on $T$.

\vspace{1\baselineskip}

In contrast to the Fermi liquid case, the Lorenz ratio $L=\dfrac{\kappa}{\sigma T}=\dfrac{\tilde{\kappa}}{\tilde{\sigma}}$ is not given by a universal constant but depends on the interaction strength $g=\dfrac{\lambda^2}{2 N}$. This is due to the infrared (IR) dependence on the integration over $\mathbf{q}$ for the optical conductivity. Because the integral is divergent, it explicitly depends on the IR cutoff which is ${\mu}/{T}$, and leads to the $g$-dependence of L. The Lorenz ratio increases with the bosonic coupling in this regime, as shown in Fig. \ref{fig:Lorentz}.

\end{document}